\documentclass[twocolumn,aps,prl,10pt,amsmath,amssymb,nofootinbib,showpacs,superscriptaddress,floatfix]{revtex4-1}
\pdfoutput=1

\DeclareFontFamily{U}{rcjhbltx}{}
\DeclareFontShape{U}{rcjhbltx}{m}{n}{<->rcjhbltx}{}
\DeclareSymbolFont{hebrewletters}{U}{rcjhbltx}{m}{n}

\usepackage{graphicx}
\usepackage{color}
\usepackage[usenames,dvipsnames]{xcolor}
\usepackage[colorlinks=true,linkcolor=Red,citecolor=Green,linktoc=page]{hyperref}
\usepackage{multirow}
\usepackage{float}
\usepackage{flushend}
\usepackage{balance}
\usepackage[varg]{txfonts}
\usepackage{ulem}
\usepackage{fancyhdr}

\begin{document}
\title{Combinatorial quantum gravity and emergent 3D quantum behaviour}

\author{C.\,A.\,Trugenberger}

\affiliation{SwissScientific Technologies SA, rue du Rhone 59, CH-1204 Geneva, Switzerland}

\begin{abstract}
We review combinatorial quantum gravity, an approach which combines Einstein's idea of dynamical geometry with Wheeler's ``it from bit" hypothesis in a model of dynamical graphs governed by the coarse Ollivier-Ricci curvature. This drives a continuous phase transition from a random to a geometric phase, due to a condensation of loops on the graph. In the 2D case, the geometric phase describes negative-curvature surfaces with two inversely related scales, an ultraviolet (UV) Planck length and an infrared (IR) radius of curvature. Below the Planck scale the random bit character survives: chunks of random bits of the Planck size describe matter particles of excitation energy given by their excess curvature. Between the Planck length and the curvature radius, the surface is smooth, with spectral and Hausdorff dimension 2; at scales larger than the curvature radius, particles see the surface as an effective Lorentzian de Sitter surface, the spectral dimension becomes 3 and the effective slow dynamics of particles, as seen by co-moving observers, emerges as quantum mechanics in Euclidean 3D space. Since the 3D distances are inherited from the underlying 2D de Sitter surface, we obtain curved trajectories around massive particles also in 3D, representing the large-scale gravity interactions. We shall thus propose that this 2D model describes a generic holographic screen relevant for real quantum gravity. 

\end{abstract}
\maketitle

\section{Introduction}

Reconciling quantum mechanics (QM) with general relativity (GR) is one of the most daunting and fundamental problems of contemporary physics. On one side, the background independence of GR is not compatible with the absolute time of QM \cite{kiefer, kuchar}; on the other side, GR is perturbatively non-renormalizable. Various routes have been proposed to cure these issues. One, is to consider GR as an effective field theory, i.e. to ditch local quantum field theory below the Planck scale in favour of a different principle, like string theory (see e.g. \cite{strings}) or abandoning local Lorentz symmetry at these small scales \cite{horava}. Another is to look for symmetry relations that reduce the physical high-energy degrees of freedom to a finite set, the asymptotic safety programme (for a review see \cite{as1, as2}). A further one is canonical quantization in terms of a different set of variables, as in loop quantum gravity (for a review see \cite{loopqg}). Finally, one can try a non-perturbative approach by formulating gravity on a ``lattice" of simplices, as in causal dynamical triangulations (CDTs) (for a review see \cite{cdt1,cdt2}), in group field theory \cite{oriti}, or by using the tensor model \cite{tensor1, tensor2, sasakura}. Finally, there are also models in which space-time is treated as a structure growing according to a causal order, as in causal set theory (for a review see \cite{causalsets}) or in the Wolfram model (for a review see \cite{gorard}). 

A different idea has been proposed by J. A. Wheeler \cite{wheeler}. This is the famed ``it from bit" hypothesis, which posits that the fundamental theory of the universe should be formulated exclusively in terms of information-theoretic binary degrees of freedom, with no reference to either time or space, which are self-synthetized concepts at large scales. Recently, this idea has been promoted to a ``it from qubit" hypothesis, which would, however, imply that quantum mechanics is more fundamental than space-time itself (for a recent review see \cite{qi}), at odds with Wheeler's original proposal. This idea is the basis for emergent quantum gravity models in which space-time emerges from entanglement of microscopic quantum degrees of freedom \cite{verlinde}. Contrary to this approach, as we will see, in the present model gravity and quantum mechanics emerge simultaneously from classical bits.

Recently, I proposed to combine Einstein's idea of dynamical geometry with Wheeler's ``it from bit" programme by formulating a purely combinatorial model of quantum gravity \cite{comb1}, in which the bits are the ``yes or no" connections between an abstract set of points, i.e. the edges of a random graph (for a review see \cite{graphrev}). The random graph can even be assembled dynamically in a phase transition, starting from completely independent Ising-like degrees of freedom \cite{first}. In analogy to GR, the Hamiltonian governing the partition function is taken as the combinatorial Ollivier-Ricci curvature of the graph \cite{olli0,olli1,olli2} (see also \cite{olli3,olli4}). The idea of applying network theory to the emergence of geometric manifolds has been also pursued in \cite{krioukov, krioukovdesitter, bianconi1, bianconi2, bianconi3}, although none of these works used a combinatorial Ricci curvature to govern the network distribution. 

GR is a theory of dynamical manifolds. We are used to associate geometry with continuum manifolds but, in principle, geometry needs only a metric. Indeed, the field of discrete geometry has advanced by leaps and bounds in recent years. The Ollivier curvature is not the only measure of discrete combinatorial curvature that has been proposed, but it is the only one that has been shown to converge to continuum Ricci curvature for geometric random graphs defined on manifolds \cite{conv1, conv2} (see \cite{tee} for its relation to Forman curvature \cite{forman}), although a recent result claims a sectional graph measure that also converges \cite{duplessis} to its continuum counterpart. The Ollivier curvature is unfortunately cumbersome to compute with, in the general case. To solve this problem, simpler approximations have been proposed \cite{klit1,klit2, klit3}. Fortunately, however, for the relevant class of random graphs, the original Ollivier measure becomes very simple. 

This is a review of the main results obtained so far in the combinatorial quantum gravity programme. We shall focus on the main ideas and concepts, referring to the original literature for technical details and derivations. The model shows a continuous phase transition from a random phase to a geometric phase as the gravitational coupling $g$ is decreased \cite{comb1, comb2}, with the emergence of manifolds at the critical point due to a condensation of loops on the graph. While the one-dimensional (1D) model is essentially exactly solvable \cite{comb3}, things become more complicated in higher dimensions. In 2D, which will be the focus of this review, geometric graphs are discretizations of Riemann surfaces of decreasing genus when the coupling $g$ is decreased, the exact ground state at $g=0$ being a torus graph. As usual in continuous phase transitions there is an emergent scale $\ell_{\rm P}$ that diverges at the critical point: in this case it describes the typical size of random graph bubbles (the disordered phase) on the manifold (the ordered phase). In the continuum limit, the emergent manifolds are negative curvature surfaces, on large scales, sprinkled by bubbles of random phase of the typical size $\ell_{\rm P}$, on small scales. Since the dominant component of diffusion in negative curvature is asymptotically equivalent to geodesic free fall on the corresponding Lorentzian de Sitter space of positive curvature \cite{eff1, eff2}, we have the dynamic emergence of coordinate time on large scales, while the small-scale random bubbles appear as particles with excitation energy given by their excess curvature. Space-time and matter are thus made of the same stuff, bits, realizing just two different phases. The (2D) big bang is a critical point at which the effective de Sitter space-time emerges. The curvature decreases simultaneously with the Planck length as more and more matter is transformed into space-time until the 2D universe becomes a flat empty surface. 
Due to the central limit theorem in negative curvature \cite{cllt, anker}, the subdominant, random component of asmyptotic diffusion sees 3 Euclidean spectral dimensions independently of the topological dimension of the manifold and its curvature; moreover, its effective dynamics coincides with (3+1)-dimensional quantum mechanics. 

Both the Hausdorff and the spectral dimension are scale-dependent. The Hausdorff dimension diverges at the Planck scale $\ell_{\rm P}$, where the random graph character starts to appear and decreases to 2 (D) on larger scales. The spectral dimension is 2 (D) on small scales and increases to 3 at large scales. Even in 2D, thus, this model of quantum gravity is different from both Liouville gravity \cite{polyakov} (for a review see \cite {seiberg}) and CDTs \cite{cdt1, cdt2}. In the final section we will suggest that it may describe the physics of a 2D holographic screen relevant for real quantum gravity.

\section{Combinatorial quantum gravity}

Combinatorial quantum gravity \cite{comb1, comb2} is formulated as a statistical partition function ${\cal Z}$ on the space of all incompressible 2D (even)-regular graphs (IERG) with $N$ vertices,
\begin{equation}
{\cal Z} = \sum_{IERG} {\rm e}^{-{1\over g} H} \ ,
\label{comqg}
\end{equation} 
where $g$ is the dimensionless coupling. Incompressible graphs are those for which short cycles do not share more than one edge, short cycles being those that matter for discrete locality on a graph,  i.e. triangles, squares and pentagons \cite{olli3}. This condition is a crucial ingredient of the model: it is the loop equivalent of the hard-core condition for bosonic point particles. As the hard-core condition prevents the infinite compressibility of Bose gases, the graph incompressibility condition prevents graph crumpling by requiring that loops can ``touch" but not ``overlap" on more than one edge. 

The incompressibility condition can be formulated as an excluded sub-graph condition \cite{comb2}, which immediately leads to an alternative interpretation. On a Riemannian manifold, the metric component $g_{ii}(x) $ defines an infinitesimal segment of length $\sqrt{g_{ii}(x)dx^i dx^i} $ at $x$ in direction $i$. The Ricci curvature then involves two derivatives of the metric. In the discrete setting of a graph, the corresponding notion of locality at a vertex involves thus distances of at most three edges. This is why the Ollivier curvature is influenced by triangles, squares and pentagons but not by longer cycles \cite{olli4}. On a continuum Riemannian manifold, the metric component $g_{ij} (x)$ at a given point $x$ defines a unique infinitesimal surface element with area $g_{ij} dx^i dk^j $ at that point. In the discrete setting, the corresponding condition requires that triangles, squares and pentagons be uniquely defined by two edges. The excluded sub-graphs are exactly those for which this condition is not satisfied. In other words, incompressible graphs are those that satisfy the necessary condition for admitting a smooth continuum limit leading to a Riemannian manifold.
It is this constraint that ultimately causes 2D combinatorial quantum gravity to be in a different universality class than Liouville quantum gravity \cite{polyakov}. Note also that some of these constraints on configuration space can alternatively be dynamically implemented by adding Kronecker-delta terms to the Hamiltonian \cite{evnin}. 

The Hamiltonian is the total Ollivier-Ricci curvature, a discrete graph equivalent of the Riemannian Einstein-Hilbert action, 
\begin{equation}
H = -{2D} \sum_{i \in G} \kappa (i) = -{2D}\sum_{i\in G} \sum_{j\sim i} \kappa (ij) \ ,
\label{deh}
\end{equation}
where we denote by $j \sim i$ the neighbour vertices $j$ to vertex $i$ in graph $G$, i.e. those connected to $i$ by one edge $(ij)$. Here $\kappa (ij)$ is the coarse Ollivier-Ricci curvature of edge $(ij)$ \cite{olli0, olli1, olli2, olli3, olli4} and $2D$ is the degree of the regular graph. 
Ricci curvature on manifolds is a measure of how much (infinitesimal) spheres around a point contract (positive Ricci curvature) or expand (negative Ricci curvature) when they are transported along a geodesic with a given tangent vector at the point under consideration. The Ollivier curvature is a discrete version of the same measure. For two vertices $i$ and $j$ it compares the Wasserstein (or earth-mover) distance $W\left( \mu_i, \mu_j \right)$ between the two uniform probability measures $\mu_{i,j}$ on the spheres around $i$ and $j$ to the distance $d(i,j)$ on the graph and is defined as
\begin{equation}
\kappa (i,j)= 1- {W\left( \mu_i, \mu_j \right) \over d(i,j)} \ .
\label{olli}
\end{equation}
The Wasserstein distance between two probability measures $\mu_i$ and $\mu_j$ on the graph is defined as
\begin{equation}
W\left( \mu_i, \mu_j \right) = {\rm inf} \sum_{i,j} \xi(i,j)d(i,j) \ ,
\label{wasser}
\end{equation}
where the infimum has to be taken over all couplings (or transference plans) $\xi(i,j)$ i.e. over all plans on how to transport a unit mass distributed according to $\mu_i$ around $i$ to the same mass distributed according to $\mu_j$ around $j$. 

While the Ollivier-Ricci curvature is cumbersome to compute in the generic case, it simplifies substantially on 
2D-regular incompressible graphs \cite{comb2}, 
\begin{equation}
\kappa(ij)= \frac{T_{ij}} {2D}-\left[1-\frac{2+T_{ij} +S_{ij}}{2D}\right]_+ - \left [1-\frac{2+T_{ij} +S_{ij} +P_{ij}}{2D}\right]_+ \ .
\label{orc}
\end{equation}
where the subscript ``+" is defined as $[\alpha]_+ = {\rm max} (0, \alpha)$ and $T_{ij}$, $S_{ij}$ and $P_{ij}$ denote the number of triangles, squares and pentagons supported on egde $(ij)$. This gives
\begin{eqnarray}
H &&= H_{\rm global} + H_{\rm local} \ ,
\nonumber \\
H_{\rm global} &&= 16 \left( {D(D-1)\over 2} N - {9\over 8} T - S- {5\over 8} P \right) \ ,
\nonumber \\
H_{\rm local} &&= \sum_{(ij) \in E_1} \left( (T_{ij} + S_{ij} ) - (2D-2) \right) 
\nonumber \\
&&+ \sum_{(ij) \in E_2} \left( (T_{ij} + S_{ij} +P_{ij}) - (2D-2) \right) \ ,
\label{globallocal}
\end{eqnarray}
where $T$, $S$ and $P$ are the total numbers of triangles, squares and pentagons on the graph and $E_1$ and $E_2$ are the ensembles of edges for which the respective summands are strictly positive. 

\section{Loop condensation, the emergence of hyperbolic manifolds}

\begin{figure}[t!]
	\includegraphics[width=9cm]{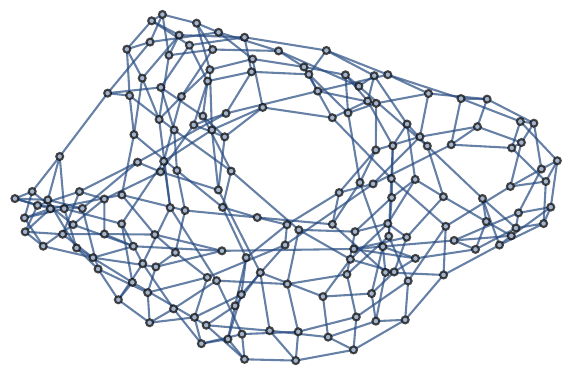}
	\vspace{-0.3cm}
	\caption{A two-torus at a small but finite value of the coupling $g$ in the geometric phase }
	\label{fig:Fig.1}
\end{figure}

The first, global term in (\ref{globallocal}) is the Hamiltonian of a matrix model. If only this term is retained, the model undergoes a first-order phase transition in which the graph decomposes into isolated, weakly interacting hypercubic complexes \cite{comb2, gorsky1, gorsky2}. If the full Ollivier-Ricci curvature is used, instead the model undergoes a continuous phase transition from a random graph phase to a geometric phase in which graphs become discretizations of manifolds \cite{comb1, comb2}. This phase transition is due to a condensation of square loops on the graph, while triangles and pentagons are suppressed and survive at best as isolated defects \cite{comb2}. As is evident from (\ref{globallocal}), the condensation of squares decreases the energy until a configuration corresponding to a regular lattice, with $S=D(D-1)N/2$ and (2D-2) squares per edge is reached. This identifies the parameter $D$ with the topological dimension of the emerged manifold. In the case $D=2$, on which we will focus from now on, the ground state is thus a torus graph. Since random graphs can be embedded without edge crossings on a Riemann surface of genus N, decreasing the coupling, in this phase, entails decreasing the genus of the emerged manifold until the torus graph at $g=0$ is reached. In Fig. \ref{fig:Fig.1} we show the example of a two-torus emerged at a small but finite value of $g$ in the geometric phase.

\begin{figure}[t!]
	\includegraphics[width=10cm]{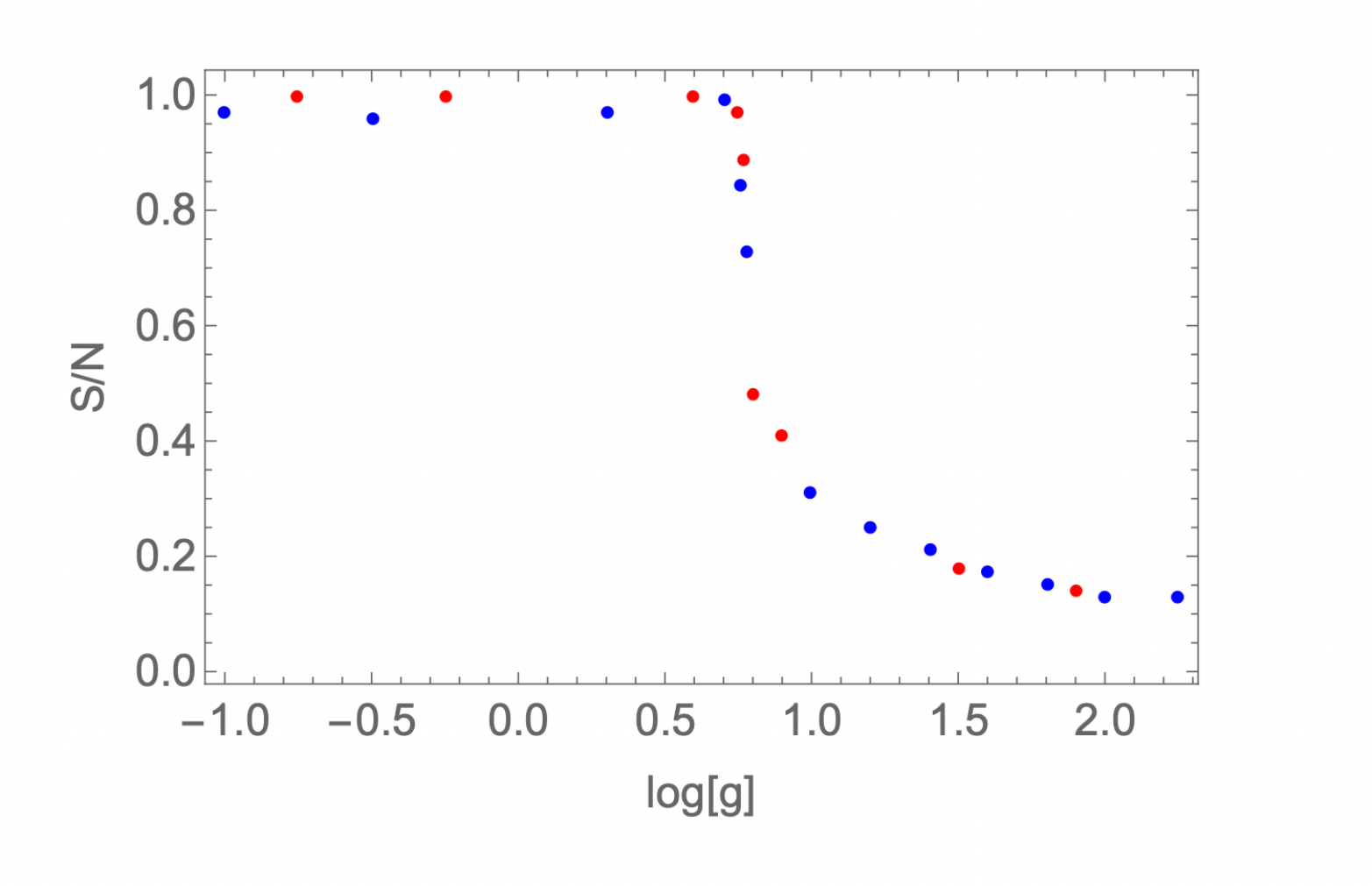}
	\vspace{-0.3cm}
	\caption{The order parameter $S/N$ for the random-to-geometric phase transition for $D=2$. }
	\label{fig:Fig.2}
\end{figure}

Random regular graphs are locally tree-like, with very sparse cycles, governed by a Poisson distribution with mean $(2D-1)^k/k$ for cycles of length $k$ \cite{oneil}. The order parameter for the transition is thus the relative number of squares $S/N$ where $N$ is the number of squares for a torus graph (for $D=2$). It is shown in Fig. \ref{fig:Fig.2} as a function of ${\rm log} (g)$. 

The red dots are obtained by starting with an exact torus graph and gradually increasing the coupling $g$; the blue dots, instead, correspond to decreasing the coupling from the random phase. Note the absence of hysteresis, as expected for a continuous phase transition. In the random phase, $S/N$ approaches $(3^4/4)/160= 0.126$ ($N=160 $ in this example), as expected from the sparse cycle distribution in this phase. As could be expected, the Monte Carlo algorithm does not always manage to reach an exact ground state with $S/N=1$ deep in the geometric phase when starting from a random graph configuration. 

\begin{figure}[t!]
	\includegraphics[width=8cm]{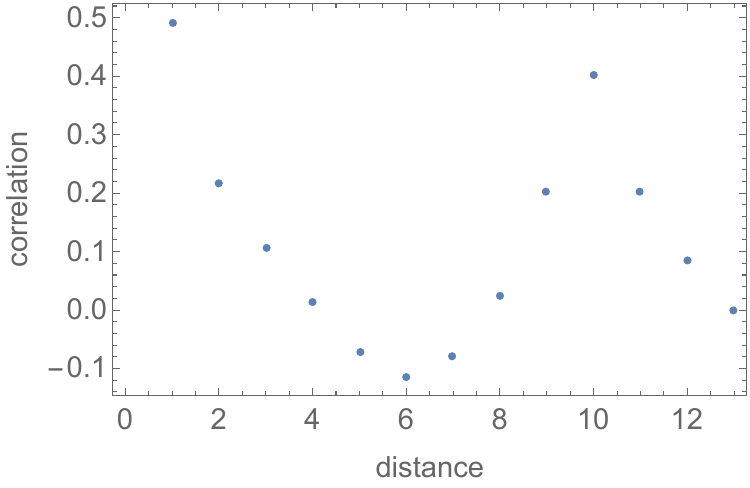}
	\vspace{-0.3cm}
	\caption{The typical behaviour of the connected correlation function between the numbers of squares at distance $d$ on the graph in the geometric phase. }
	\label{fig:Fig.3}
\end{figure}

As always in the case of continuous phase transitions, there appears a length scale $\xi(g)$ diverging at the critical point. Let us consider the connected correlation function
\begin{equation}
c (d, g) = {\langle \left( S_i- \bar S \right ) \left( S_j- \bar S \right ) \rangle \over \sigma^2 (S)} \ ,
\label{correlation}
\end{equation}
between the number of squares $S_i$ and $S_j$ based on vertices $i$ and $j$ at fixed graph distance $d$, where $\sigma^2 (S)$ denotes the variance (over the vertices) of the number of squares. The typical behaviour of this correlation in the geometric phase, as a function of $d$ (at fixed $g$) is shown in Fig. \ref{fig:Fig.3}.

Note the intermediate domain in which correlations turn negative. This is because there is no correlation between squares on these distances but we are in the geometric phase, in which the average square number $\bar S$ is large. On these intermediate distance the graph behaves like a random graph. Correlations then suddenly pick up at the distance $\xi(g) =10$, before decaying again to zero on even larger distances. In this case, the approach to zero is not due to absence of correlations but, rather, to the fact that each vertex has the same number of squares approaching $\bar S$. This is the geometric manifold phase on large scales. 

The graphs in the geometric phase behave thus as geometric manifolds on distances larger than $\xi(g)$ and as random graphs below this scale. This has an important consequence. In random graphs, distances scale as ${\rm log} N_\xi$ \cite{graphrev}, where $N_\xi = N^{\alpha(g)} $ is the typical number of vertices in them, with $0 \le \alpha(g) \le 1$, whereas they scale linearly with $N$ in the geometric phase. The Hausdorff dimension is thus a scale-dependent quantity that diverges on the scale $\xi(g)$ and decreases to 2 (D in general) on larger scales. We can now take the continuum limit by introducing a fixed length $\ell$ for the graph edges and letting simultaneously $\ell \to 0$ and $N \to \infty$ so that $\ell_{\rm P} = \alpha(g) \ell {\rm log} N$ is fixed. We obtain so an infinite surface (manifold in general), sprinkled on the scale $\ell_{\rm P}$ by bubbles of random phase (the disordered phase). 

\begin{figure}[t!]
	\includegraphics[width=8cm]{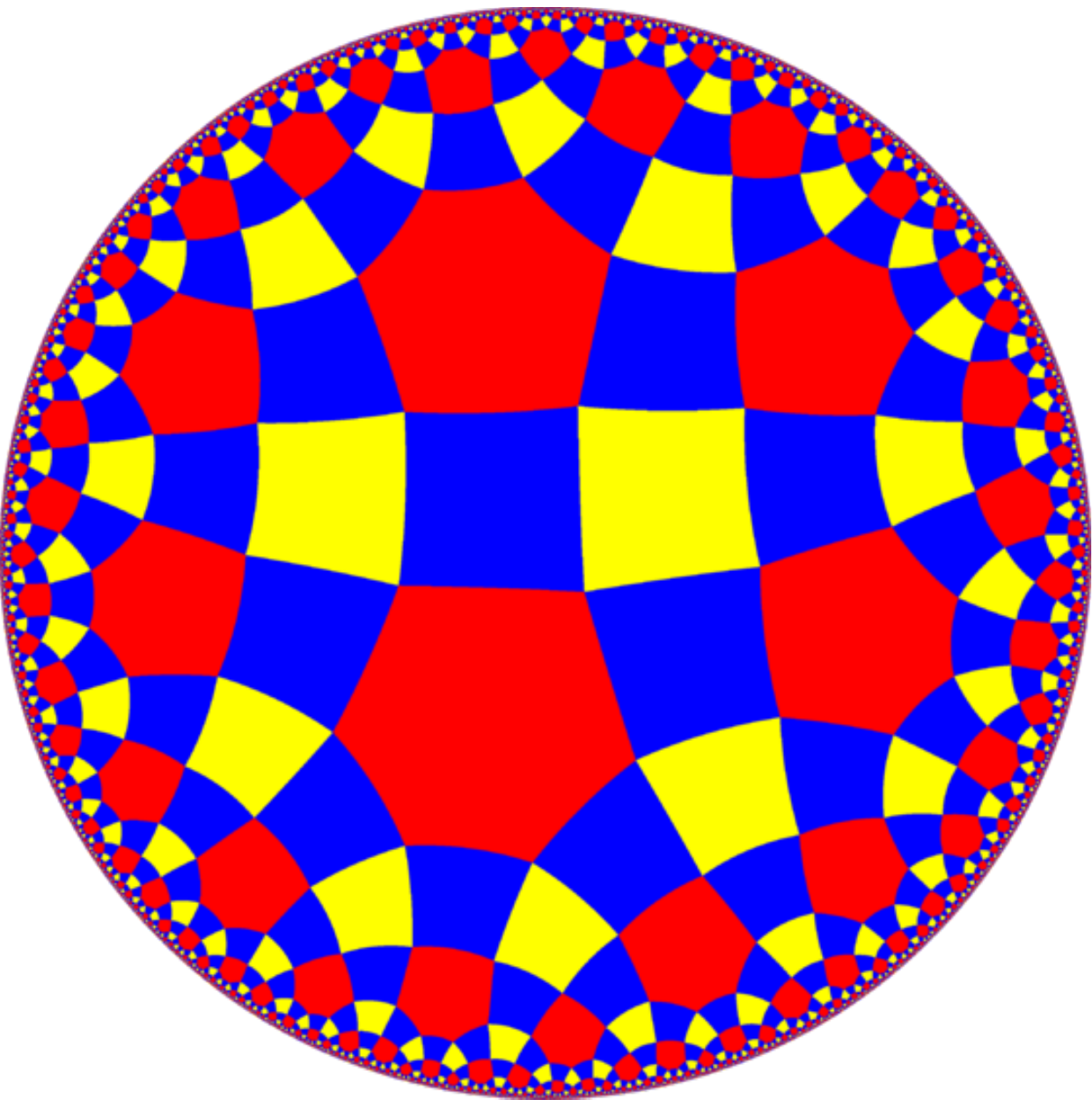}
	\vspace{-0.3cm}
	\caption{The 2D tiling ``geometrizing" the 3-square-per-vertex infinite graph via its 1-skeletons.}
	\label{fig:Fig.4}
\end{figure}

To establish what is the character of the emerging surfaces on large scales, let us for the moment forget the microscopic structure and consider only infinite graphs with the same number of squares per vertex, say 3, for example. In order to ``geometrize" the graphs, we must turn these purely combinatorial objects into 1-skeletons of topological spaces by assigning a fixed length $\ell$ to the edges, as discussed above. When this is done, the 2-cell embeddings become 1-skeletons of tilings of the sought after continuum surface. In general this is not easy. In 2D, however, it can be done \cite{datta}. We show such a tiling for the 2D case of 3 squares per vertex in Figure \ref{fig:Fig.4}.

The surface defined by this tiling is the hyperbolic plane. On large scales, the emerged manifolds are thus surfaces of negative curvature, so called Cartan-Hadamard manifolds (for a review see \cite{shiga}) in general or hyperbolic manifolds (for a review see \cite{anderson}) when no defects are present and the negative curvature is constant. The spectral dimension fits confirm that the geometric phase manifolds on intermediate scales, just above $\ell_{\rm P}$ are indeed hyperbolic surfaces \cite{eff1}. We shall return below to the very large scales.

Let us now return to the small scale bubbles of randomness. Since the Hamiltonian of the model is the negative Ollivier curvature and random graphs have the lowest possible Ollivier curvature, these bubbles have an excitation energy corresponding to minus the excess negative curvature with respect to the geometric manifold at large scales. Note also that the UV scale $\ell_{\rm P}$ is inversely related to the IR scale given by the radius of curvature of the geometric phase. When $\ell_{\rm P}$ vanishes for $g \to 0$, the radius of curvature diverges and the surface becomes flat. For $g \to g_{\rm crit}^-$, the radius of curvature diverges, while $\ell_{\rm P}$ increases until the whole space fits into a dot of size $\ell_{\rm P}$ within which distances scale logarithmically.

\section{Emergence of time and effective de Sitter space}

So far so good as far as statics is concerned. As in any statistical mechanics model, however, there will be fluctuations about the free energy minima. For this, however we need a universal time, which we shall call $t$. This has nothing to do with the coordinate time of GR, however, we need only an ordinal quantity that defines ``before vs. after". But we have already such a quantity in the model: it is the gravitational coupling $g$ itself, we do not need to introduce a new concept, it is sufficient to take $\beta_g = 1/g$ as universal time $t$. This choice is natural, since it leads immediately to a cosmological model, as we will mention below.  

As a consequence of these fluctuations, the randomness bubbles will undergo small changes in shape, size and position. A full treatment of the shape and size fluctuations is beyond the scope of the present review. Here we shall consider the bubbles as point-like objects and focus only on their positions. In this limit, we are left with Brownian motion, which captures exactly the scattering of the point particles on background fluctuations. Brownian motion, however, is sensitive to the geometry of the fluctuating background: in the case of negative curvature it is anomalous and becomes asymptotically ballistic. 

Let us consider a constant negative curvature surface as the upper sheet of a two-sheeted hyperboloid embedded in 3D Minkwoski space with metric $(+1,+1, -1)$ as
\begin{equation}
{\bf x} = \begin{pmatrix} {1\over H} {\rm sinh} Hz \ {\rm cos} \theta \\ {1\over H} {\rm sinh} Hz \ {\rm sin} \theta \\ {1\over H} {\rm cosh} Hz \end{pmatrix} \ ,
\label{twosheets}
\end{equation}
The metric is $ds^2 = dz^2 + G^2(z) d\theta^2 $, with $G(z) = {{\rm sinh} Hz / H}$
and $-H^2$ the curvature. 

Brownian motion on this manifold decomposes in two independent radial and angular Brownian motions $(z_t, \theta_t)$ \cite{prat, kendall, hsu} (for a review see \cite{hsubrief, hsubook, arnaudon}). For $Hz \gg 1$ the radial equation reduces to 
\begin{equation}
z_t= {H\over 2} t + R_t \ ,
\label{rad}
\end{equation}
up to exponentially small corrections $O\left({\rm exp}(-Hz)\right)$. Here, $R_t $ is a sub-dominant random component such that
$R_t/t = O\left( t^{-1/2} \right)$ for large $t$. Note that, for simplicity of presentation, a diffusion coefficient ${\cal D}$ is understood as absorbed in the universal time $t$, which has thus dimension $[length]^2$. The parameter $\tilde t =t/{\cal D}$ is thus measured in seconds and $v=H{\cal D}$ is a velocity. This equation means that, asymptotically, the manifold coordinate $z$ can be identified with the absolute time \cite{eff1, eff2}. This is the dynamic emergence of coordinate time in this region of the manifold. We shall discuss the the sub-dominant random component below.

The angular Brownian motion converges to a limiting angle $\theta_\infty$. Its quadratic variation is given by \cite{hsubrief, arnaudon} 
\begin{equation} 
\langle \Delta \theta_t ^2 \rangle = \int_0^t ds {1\over G\left( z_s \right)^2} \ ,
\label{quadvartot}
\end{equation} 
which, near the limiting angle becomes
\begin{equation} 
\langle \Delta \left(\theta_\infty- \theta_t \right)^2  \rangle= \int_t^\infty ds {1\over G\left( z_s \right)^2} \ .
\label{quadvarinf}
\end{equation} 
If we consider large values of $t$ we can substitute $z_s$ with its limiting value (\ref{rad}). Then, using the above value of $G(z)$ we obtain the dominant contribution to the angular Brownian motion as
\begin{equation}
\langle \Delta \left(\theta_\infty-\theta_z \right)^2 \rangle = 4 \ {\rm e}^{-2Hz} \ .
\label{limquadmot}
\end{equation}
This is the exact inverse of the usual Brownian motion, whose quadratic variation is quadratic in $t$ (and thus ballistic) at short times, below the typical scattering time and linear in $t$ (and thus diffusive) at large times, when scatterings become frequent. The convergence to a limiting angle has the consequence that scatterings become less and less frequent asymptotically until they stop altogether and motion becomes ballistic in the time parameter $\tau = 2 \ {\rm exp} (-Hz)$, with the distribution probability $u(z, \theta)$ satisfying 
\begin{equation}
\left( {\partial^2 \over \partial \tau^2} - {\partial^2 \over \partial \theta^2}  \right) u = 0 \ ,
\label{wave1}
\end{equation} 
which is equivalent to
\begin{equation}
\left( {\partial^2 \over \partial z^2} + H {\partial \over \partial z} - 4H^2 {\rm e}^{-2Hz} {\partial^2 \over \partial \theta^2} \right) u(z, \theta) = 0 \ .
\label{kleingordoncurved}
\end{equation}
This is the wave equation on Lorentzian de Sitter space-time \cite{strominger, cosconst} of positive curvature $+H^2$, obtained from (\ref{twosheets}) by the substitution ${\rm cosh}(Hz) \leftrightarrow {\rm sinh}(Hz)$. Note that the diffusion coefficient falls out of this equation when written in terms of coordinate time. Up to exponentially small correction and the sub-dominant random component to be discussed below, Brownian motion in constant negative curvature is equivalent to geodesic free fall on de Sitter space-time. The same picture applies to the flat slicing of both manifolds \cite{eff1, eff2}. Lorentzian de Sitter space-time appears thus as an effective, rather than fundamental, description of the fluctuating manifold in the region $Hz \gg 1$ (note that the radial coordinate $z$ has an absolute physical meaning because the microscopic defects break the homogeneity of pure hyperbolic space). The prescription for going from the fundamental to the effective description of the manifold is a simultaneous change of sign of the time component of the metric and the curvature. For flat space operators, this amounts to the usual Wick rotation which, however is here induced dynamically.
Of course, in the effective description, the randomness bubbles have a natural interpretation as matter particles of the Planck size $\ell_{\rm P}$, with an intrinsic rest energy proportional to the excess (positive) curvature with respect to the de Sitter background. In this model, thus space-time and matter are just two different phases of the same stuff, bits. In the course of absolute time, i.e. when the coupling $g$ decreases, the whole space-time emerges from a random graph dot of the Planck size and its curvature decreases when more and more matter is transformed into space until a flat, empty manifold is reached. In the course of this evolution, the Lorentzian effective region is pushed out more and more until it vanishes completely.

\section{Spectral dimension 3 and quantum behaviour} 

In general, diffusion processes probe the intrinsic geometry of a manifold via the return probability kernel $K(t) =  {\rm Tr} \ {\rm exp}(t \Delta)$. The quantity $d_{\rm s} = -2 d{\rm ln}\ K(t)/d{\rm ln}\ t$ is called the spectral dimension and measures the effective number of dimensions available to a random walker on the scale reached by time $t$, essentially on small scales for $t\to 0$ and on very large scales for $t \to \infty$. In the mathematical literature the latter is called ``geometry at infinity". However, this does not work in the present case
because the Laplacian on a constant negative curvature manifold has a spectral gap
\begin{equation}
\lambda_0 = - {\rm lim}_{t\to \infty} {{\rm ln}\ K(t) \over t} = {(D-1)^2H^2\over 4}  \ ,
\label{spegap}
\end{equation} 
representing the bottom of the spectrum of the positive operator $-\Delta$. As a consequence, the return probabilities $K(t)$ are dominated by an exponential behaviour at large $t$  \cite{davies, grigorian}
\begin{equation}
K(t) \asymp {\left( 1+H^2 t \right)^{(D-3)/2} \over (H^2t)^{D/2}} {\rm e}^{ -{(D-1)^2 \over 4} H^2 t} \ ,
\label{returnCH}
\end{equation}
which gives a spectral dimension linearly diverging in time. 

The large-scale geometry, however, can be probed by the sub-dominant random component of Brownian motion \cite{anker}. This is called the infinite Brownian loop \cite{anker}, and is the infinite-$\Theta$ limit of the Brownian bridge $BB(t,\Theta)$, which is the Brownian motion $B(t)$ constrained to come back to the origin $x=0$ for $t = \Theta$. 
Let us decompose the Brownian process $B(t)$ as
\begin{equation}
B(t) =  {t\over \Theta} B(\Theta) + BB(t,\Theta) \ .
\label{bridge}
\end{equation} 
When Brownian motion is not asymptotically ballistic, the infinite Brownian loop is the Brownian process itself. Otherwise, it represents the subdominant random component after the dominant deterministic component has been subtracted. The corresponding return probabilities are given by (\ref{returnCH}) without the dominant exponential factor and give the spectral dimension function
\begin{equation} 
d_{\rm s} (t)  = D- (D-3) \left( {H^2 t\over 1+H^2 t} \right) \ .
\label{modspe}
\end{equation}
Thus, the spectral dimension on scales smaller than $1/H$ is $D$ while the large-scale spectral dimension above distances $1/H$ becomes 3, independently of $D$. In the mathematical literature this is called the  ``pseudo-dimension", or ``dimension at infinity" of the constant negative curvature manifold \cite{anker}. The large-scale spectral dimension 3 is not confined to a constant negative curvature manifold but is valid for any manifold with strictly non-positive curvature and is a consequence of the central limit theorem in negative curvature \cite{cllt}. 

The infinite Brownian loop is the diffusion process of a particle as seen by an observer co-moving with the dominant ballistic flow. In mathematical terms it is the relativized $\varphi_0$-process $f_{\rm BL}\varphi_0$, where $\varphi_0$ is the symmetric ground state of the Laplacian corresponding to the spectral gap, $\left( \Delta + \lambda_0 \right) \varphi_0 = 0$ \cite{anker}. It is generated by 
\begin{equation}
\tilde \Delta \left( f_{\rm BL} \right) = {1\over \varphi_0} \Delta \left( f_{\rm BL} \varphi_0 \right) + \lambda_0 f_{\rm BL} = 
\Delta f_{\rm BL} + 2 \nabla {\rm ln} \varphi_0 \cdot \nabla f_{\rm BL} \ .
\label{relat}
\end{equation}
Using the known ground state $\varphi_0$ \cite{davies} one can see that the second term on the right-hand side of this expression transforms (asmyptotically) the negative-curvature Laplacian on a D-dimensional manifold into the flat Laplacian in 3D, so that the uniform continuous estimate of the kernel of the infinite Brownian loop is given by \cite{davies}:
\begin{equation}
K(t, \rho) \asymp t^{-{3\over 2}} {\rm e^{-{\rho^2 \over 4t}}} \ ,
\label{iblkernel}
\end{equation}
where $\rho$ is the hyperbolic distance. This is the isotropic heat kernel on a 3D Euclidean manifold with distance norm $\rho$ ``inherited" from the 2D hyperbolic distance \cite{anker}. The corresponding kernel in the effective Lorentzian description is then
\begin{equation}
K_{\rm eff} (\tilde t, \rho) \asymp \left( {m\over i\hbar \tilde t} \right)^{3\over 2} \  {\rm e^{-{m \rho^2 \over 2i\hbar  \tilde t}}} \ ,
\label{qm}
\end{equation}
where we have used the previously introduced time $\tilde t$ measured in seconds. This is the 3D Schrödinger propagator for a particle of mass $m=\hbar H/ 2v$ and velocity $v=H{\cal D}$. At large scales, the effective description of slow processes, as seen by geodesically free-falling observers on the effective de Sitter surface, is 3D quantum mechanics with distances inherited from the fundamental 2D hyperbolic surface. 

\section{Concluding proposal}
We have discussed the emergence of hyperbolic surfaces from regular random graphs at a critical point corresponding to the condensation of 4-cycles (squares) on the graph. While there are concrete hints that the overall picture is not much different, the treatment of $D>2 $ manifolds is beyond the scope of presently available numerical power. Moreover, the geometrization proof is also not established for $D>2$. Here, however, we would like to suggest that the $D=2$ case is actually sufficient as a model of quantum gravity.  

The idea is that the hyperbolic surface constitutes a generic holographic screen, which is neither a boundary nor a horizon, as suggested in \cite{bousso}. At short scales, smaller than its curvature radius, we see the holographic screen, with spectral and Hausdorff dimension 2, exactly as in causal dynamical triangulations and in Horava-Lifshitz gravity \cite{cdt1, horava}. At these high energies, the dynamics of particles is diffusion on the screen. 
At scales larger than the screen curvature radius, slow particles (massive ones, of course, I have not discussed wave propagation here) see (3+1)-dimensional Galilean space-time, with quantum mechanics governing particle dynamics. The 3D distances, however, are inherited from the screen hyperbolic distances. Therefore the (3+1)-dimensional ``spectral space-time" naturally inherits an isometry group $SO(1,3)$, so that the the isometry group $SO(1,2)$ of the screen is recovered when the holographic coordinate is frozen and the Galilean transformation are the $v/c \ll 1$ limit of slow particles. It is thus natural to identify this ``spectral space-time" at large scales with our familiar (3+1)-dimensional Minkowski space-time. Note also that, since particles are Planck-scale lumps of excess curvature, the screen geodesics are curved around them and this curvature is inherited in the 3D space. This inherited curvature represents the large-scale gravity interactions.

To our best knowledge, this is the only model where gravity and quantum mechanics emerge together at large scales from a more fundamental statistical model of information bits, as suggested originally by Wheeler \cite{wheeler}. As such, it might prove of importance for the resolution of the quantum gravity puzzle. Of course, here we have reviewed only the fundamental aspects; the cosmological and astrophysical implications remain to be studied and will be the crucial determinants of the relevance of the model for the real universe.


\begin{thebibliography}{10}
	
\bibitem{kiefer} C. Kiefer, Quantum Gravity, Oxford University Press, Oxford (2007). 

\bibitem{kuchar} K. V. Kuchar, Time and interpretations of quantum gravity, {\it Int. J. Mod. Phys.} {\bf D20} 3 (2011). 

\bibitem{strings} J. Polchinski, String Theory, Cambridge University Press, Cambridge (1998). 

\bibitem{horava} P. Horava, Quantum gravity at a Lifshitz point, {\it Phys. Rev. } {\bf D79} 084008 (2009). 

\bibitem{as1} A. Eichhorn, Asymptotically safe gravity, proceedings of the 57th Course of the Erice International School of Subnuclear Physics, "In search for the unexpected", June 2019, arXiv: 2003.00044, (2019). 

\bibitem{as2} A. Eichhorn, The microscopic structure of quantum space-time and matter from a renormalization group perspective, {\it Nat. Phys.} {\bf 19} 1527-1529 (2023). 

\bibitem{loopqg} A. Ashtekar and E. Bianchi, A short review of loop quantum gravity, {\it Rep. Prog. Phys} {\bf 84} 042001 (2021). 

\bibitem{cdt1} 
J Ambjorn, J. G\"orlich, J. Jurkiewicz and R. Loll, Nonperturbative Quantum Gravity, {\it Phys. Rep. } {\bf 519} 127-210 (2012) 

\bibitem{cdt2} R. Loll, Quantum Gravity from Causal Dynamical Triangulations: A Review, {\it Classical and Quantum Gravity}  {\bf 37} 013002 (2019). 

\bibitem{oriti} D. Oriti, The microscopic dynamics of quantum space as a group field theory. In: Foundations of space and time: reflections on quantum gravity. Aug 10–14 2009; Cape Town, South Africa; Cambridge University Press  257 (2011). 

\bibitem{tensor1} J. Ambjorn, B. Durhuus and T. Jonsson, Three-dimensional simplicial quantum gravity and generalized matrix models, {\it Mod. Phys. Lett.} {\bf A06} 1133-1146 (1991).

\bibitem{tensor2} N. Sasakura, Tensor model for gravity and orientability of manifold, {\it Mod. Phys. Lett.} {\bf A06} 2613-2624 (1991). 

\bibitem{sasakura} N. Sasakura, Phase profile of the wave function of canonical tensor model and emergence of large spacetimes, {\it Int. Jour. Mod. Phys.} {\bf 36} 2150222 (2021). 

\bibitem{causalsets} S. Surya, The causal set approach to quantum gravity, {\it Living Reviews in Relativity} {\bf 22:5} (2019). 

\bibitem{gorard} J. Gorard, Some Relativistic and Gravitational Properties of the Wolfram Model, {\it Complex Systems} {\bf 29} 599-654 (2020). 

\bibitem{wheeler} J. A. Wheeler, Information, physics, quantum: the search for links, Proceedings of the III international symposium on the foundations of quantum mechanics, 354-358, Tokyo (1989) . 

\bibitem{qi} X.-L. Qi, Does gravity come from quantum information?, {\it Nat. Phys.} {\bf 14} 984-987 (2018). 

\bibitem{verlinde} E. Verlinde, Emergent gravity and the dark universe, {\it SciPost Phys.} {\bf 2} 016 (2017). 

\bibitem{comb1} C. A. Trugenberger, Combinatorial quantum gravity: geometry from random bits,  {\it JHEP} {\bf 09} 045 (2017).

\bibitem{graphrev}R. Albert and L. Barabasi, Statistical mechanics of complex networks, {\it Rev. Mod. Phys.} {\bf 74} (2002) 47. 

\bibitem{first} C. A. Trugenberger, Quantum gravity as an information network self-organization of a 4D universe,
 {\it Phys. Rev.} {\bf D92} (2015) 084014.

\bibitem{olli0} Y. Ollivier, Ricci curvature of metric spaces, {\it C. R. Math. Acad. Sci. Paris} {\bf 345} 643-646 (2007). 

\bibitem{olli1}Y. Ollivier, Ricci curvature of Markov chains in metric spaces, {\it J. Funct. Anal.} {\bf 256} (2009) 810; 

\bibitem{olli2} Y. Ollivier, A survey of Ricci curvature fo metric spaces and Markov chains, {\it Adv. Stud. Pure Math.} {\bf 57} (2010) 343-381 (2010). 

\bibitem{olli3}Y. Linn, L. Lu and S. T. Yau, Ricci curvature of graphs, {\it Tohoku Math. J.} {\bf 63} (2011) 605-627. 

\bibitem{olli4} J. Jost and S. Liu, Ollivier's Ricci curvature, local clustering and curvature dimension inequalities on graphs, {\it Discrete Comput. Geom.} {\bf 51} (2014) 300-322. 

\bibitem{krioukov} D. Krioukov, Clustering implies geometry in networtks, {\it Phys. Rev. Lett.} {\bf 116} (2016) 208302. 

\bibitem{krioukovdesitter} D. Krioukov, M. Kitsak, R. S. Sinkovits, D. Rideout, D. Meyer and M. Boguna, Network Cosmology, {\it Scientific Reports} {\bf 2} 793 (2012). 

\bibitem{bianconi1} G. Bianconi and C. Rahmede, Complex quantum network manifolds in dimension d > 2 are scale-free, {\it Sci. Rep.} {\bf 10} 13979 (2015). 

\bibitem{bianconi2} G. Bianconi and C. Rahmede, Network geometry with flavor: from complexity to quantum geometry, {\it Phys. Rev.} {\bf E93} 032315 (2016). 

\bibitem{bianconi3} G. Bianconi and C. Rahmede, Emergent hyperbolic network geometry, {\it Sci. Rep.} {\ bf 7} 41974 (2017). 

\bibitem{conv1} P. van der Hoorn , W. J. Cunningham, G. Lippner, C. A. Trugenberger and D. Krioukov, Ollivier-Ricci curvature convergence in random geometric graphs, {\it Phys. Rev. Res.} {\bf 3} 013211 (2021). 

\bibitem{conv2} C. Kelly, C. A. Trugenberger and F. Biancalana, Convergence of combinatorial gravity, {\it Phys. Rev. } {\bf D105} 124002 (2022). 

\bibitem{tee} P. Tee and C. A. Trugenberger, Enhanced Forman curvature and its relation to Ollivier curvature, {\it EPL} {\bf 133} 60006 (2021). 

\bibitem{forman} R. Forman, Combinatorial Morse theory, {\it Int. J. Math} {\bf 13} 333-368  (2002).

\bibitem{duplessis} J. F. DuPlessis and X. D. Arsiwalla, A cosine rule-based discrete sectional curvature for graphs, {\it Journal of Complex Networks} {\bf 4} 1 (2023). 

\bibitem{klit1} N. Klitgaard and R. Loll, Introducing Quantum Ricci Curvature, {\it Phys. Rev.}  {\bf D97} 046008 (2018). 

\bibitem{klit2} N. Klitgaard and R. Loll, Implementing Quantum Ricci Curvature, {\it Phys. Rev.} {\bf D97} 106017 (2018). 

\bibitem{klit3} N. Klitgaard and R. Loll, How round is the quantum de Sitter universe? {\it Eur. Phys. J.} {\bf 80} 990 (2020). 

\bibitem{comb2} C. Kelly, C. A. Trugenberger, and F. Biancalana, Self-Assembly of Geometric Space from Random Graphs, {\it Classical and Quantum Gravity} {\bf 36} 125012 (2019). 

\bibitem{comb3} C. Kelly, C. A. Trugenberger and F. Biancalana, Emergence of the circle in a statistical model of random cubic graphs, {\it Classical and Quantum Gravity} {\bf 38} 075008 (2021). 

\bibitem{eff1} C. A. Trugenberger, Emergent time, cosmological constant and boundary dimension at infinity in combinatorial quantum gravity, {\it JHEP} {\bf 04} 019 (2022). 

\bibitem{eff2} C. A. Trugenberger , Effective de Sitter space, quantum behaviour and large-scale spectral dimension (3+1), {\it JHEP} {\bf 03} 186 (2023). 



\bibitem{cllt} F. Ledrappier and S. Lim, Local limit theorem in negative curvature, arXiv:1503.04156.

\bibitem{anker} J.-P. Anker, P. Bougerol and T. Jeulin, The infinite Brownian loop on a symmetric space, {\it Rev. Mat. Iberoamericana} {\bf 18} 41-97 (2002). 

\bibitem{polyakov} A. Polyakov, Quantum gravity in two dimensions, {\it Mod. Phys. Lett.} {\bf A02} 893-898 (1987). 

\bibitem{seiberg} N. Seiberg, Notes on quantum Liouville theory and quantum gravity, {\it Rev. Mod. Phys.} {\bf 102} 319-349 (1990). 

\bibitem{evnin} P. Akara-Pipattana, T. Chotibut and O. Evnin, The birth of geometry in exponential random graphs, {\it J. Phys. } {\bf A54} 425001 (2021). 

\bibitem{gorsky1}A. Gorsky and O. Valba, Interacting thermofield doubles and critical behaviour in random regular graphs, {\it Phys. Rev.} {\bf D103} 106013 (2021). 

\bibitem{gorsky2} A. Gorsky, V. Kazakov, F. Levkovich-Maslyuk and V. Mishnyakov, A flow in the forest, {\it JHEP} {\bf 03} 067 (2023). 

\bibitem{oneil}P. E. O’Neil, Asymptotics in random (0,1)-matrices, {\it Bull. Am. Math. Soc.} {\bf 75} 1276 (1969). 

\bibitem{datta} B. Datta and S. Gupta, Semi-regular tilings of the hyperbolic plane, {\it Discrete Comput. Geom.}  {\bf 65} 531 (2019).

\bibitem{shiga} K. Shiga, Hadamard manifolds, in ``Geometry of geodesics and related topics", {\it Advanced Studies in Pure Mathematics} {\bf 3} 239-281 (1984). 

\bibitem{anderson} J. W. Anderson, Hyperbolic geometry, Springer-Verlag, Berlin (2005). 

\bibitem{prat} J.-J. Prat. Etude asymptotique et convergence angulaire du mouvement brownien sur une vari\'et\'e \`a courbure n\'egative. {\it C. R. Acad. Sci. Paris S\'er. A-B} {\bf 280(22):Aiii} A1539-A1542 (1975). 

\bibitem{kendall} W. S. Kendall, Brownian motion on 2-dimensional manifolds of negative curvature. {\it Trans. Amer. Math. Soc.} {\bf 275} 751-760 (1983). 

\bibitem{hsu} P. Hsu and W. S. Kendall, Limiting angle of Brownian motion in certain two-dimensional Cartan-Hadamard manifolds {\it Annales de la facult\'e des Sciences de Toulouse} {\bf 1} 169-186 (1982). 

\bibitem{hsubrief} E. P. Hsu, A brief introduction to Brownian motion on a Riemann manifold, Summer School in Kyushu (2008). 

\bibitem{hsubook} E. P. Hsu, Stochastic analysis on manifolds, {\it Graduate studies in mathematics} {\bf 38}, Providence (RI) (2002). 

\bibitem{arnaudon}M.Arnaudon and A. Thalmeier, Brownian motion and negative curvature, {\it Progress in Probability} {\bf 64} 145-163 (2011). 

\bibitem{strominger} M. Spradlin, A. Strominger and A. Volovich, Les Houches lectures on de Sitter space, arXiv:hep-th/0110007. 

\bibitem{cosconst} T. Padmanabhan, Cosmological constant-the weight of the vacuum, {\it Phys. Rept.} {\bf 380} 235-320 (2003). 

\bibitem{davies} E. B. Davies and N. Mandouvalos, Heat Kernel bounds on hyperbolic space and Kleinian groups, {\it Proc. London Math. Soc. (3) } {\bf 52} 182-208 (1988).

\bibitem{grigorian} A. Grigor\' yan, Estimates of heat kernels on Riemannian manifolds, in ``Spectral Theory and Geometry", Cambridge University Press, Cambridge (2010). 

\bibitem{bousso} R. Bousso, The holographic principle for general backgrounds, {\it Class. Quant. Gravity} {\bf 17} 997 (2000). 


































	
	
\end{thebibliography}
\end{document}